\documentstyle[prl,aps,floats,epsf]{revtex}
\tighten

\epsfclipon

\begin{document}

\twocolumn[\hsize\textwidth\columnwidth\hsize\csname
@twocolumnfalse\endcsname

\title{Resonance-assisted tunneling in near-integrable systems}

\author{Olivier Brodier, Peter Schlagheck, and Denis Ullmo}

\address{
Laboratoire de Physique Th\'eorique et Mod\`eles Statistiques (LPTMS), 
91405 Orsay Cedex, France}

\maketitle

\begin{abstract}

Dynamical  tunneling between symmetry  related invariant tori is studied
in the near-integrable regime.   Using the kicked Harper model as
an illustration,  we show that the  exponential decay of  the 
wave functions  in the  classically  forbidden region  is  modified due  to
coupling processes  that are  mediated by classical  resonances.  This
mechanism leads  to a substantial  deviation of the  splitting between
quasi-degenerate eigenvalues from the purely exponential decrease with
$1/\hbar$ obtained for the  integrable system.  A simple 
semiclassical framework,  which  takes into  account  the  effect  of the  
resonance substructure on  the KAM tori, allows to quantitatively  
reproduce the behavior of the eigenvalue splittings.
\end{abstract}


]
\bigskip

\narrowtext

Despite  its  genuinely   quantal  character,  tunneling  is  strongly
influenced, if not entirely governed,  by the structure of the underlying
classical  phase  space \cite{Cre98}. Changing  the  dynamics  from 
integrable  to
chaotic substantially modifies the  tunnel coupling between two wells,
not only  if the chaos affects  the classical motion  within the wells
(as considered in \cite{CreWhe96PRL}),  but also if the destruction of
invariant  tori is  entirely  restricted to  phase  space domains  far
inside the classically forbidden region.

Consider,  for  instance, a  classical  system  exhibiting,  due to  a
discrete  symmetry,  two   congruent  but  separate  regular  regions.
Semiclassical  EBK quantization  enables one  to construct  ``quasi-modes''
within  each of  these regions  (i.e., wave  functions  fulfilling the
Schr\"odinger  equation to all powers of  $\hbar$), with  exactly the
same energy  for any pair  of symmetry related quantized  tori.  This
degeneracy  becomes eventually  lifted by  the  non-classical coupling
between  the   quasi-modes,  which  selects  the   symmetric  and  the
antisymmetric   linear  combination  of   these  states   as  ``true''
eigenstates of the quantum system.

This ``dynamical tunneling''  process \cite{DavHel81JCP} arises in both
integrable   and  non-integrable   systems,  but   with  substantially
different effectiveness.  Introducing  an appreciable chaotic layer in
between  the two  regular  regions significantly  enhances the  tunnel
coupling  as  compared to  the  integrable  case  and induces  a  huge
sensitivity  of the coupling  with respect  to variations  of external
parameters \cite{LinBal90PRL,BohTomUll93PR,TomUll94PRE}.
This  phenomenon  was  successfully explained by  the interaction of the 
regular  quasi-modes with quantum states living semiclassically within the 
chaotic domain, which, due to their  delocalized  nature, assist  at  the  
connection between  these quasi-modes 
\cite{BohTomUll93PR,TomUll94PRE,DorFri95PRL}. 
Based on this picture, random matrix descriptions of the chaotic part  
of phase space were shown to reproduce the statistical properties of the 
tunneling rates \cite{TomUll94PRE,DorFri95PRL,LeyUll96JPA,ZakDelBuc98PRE}.
Recent wave chaos experiments on optical \cite{NoeSto97Nat} and microwave 
cavities \cite{DemO00PRL} confirm the relevance of chaos for the tunnel
coupling in nonintegrable systems.
For the classically  forbidden  component of  the underlying coupling 
process, however, namely the continuation of the  wave function from the 
quantized torus to the chaos border, a general semiclassical theory is still 
lacking.

In the present contribution, we shall not tackle this problem directly
for strongly mixed dynamics,  but consider a simpler situation, namely
the  nearly  integrable regime  for  which  chaos  is not  appreciably
developed.  Yet,  we shall see  that, despite a  seemingly ``regular''
phase  space,  the  tunnel decay  of the  wave  function  and  the
associated coupling rates are  non-trivial and cannot be reproduced by
an  integrable approximation  of the  dynamics.  To  explain  this, we
shall emphasize the role of {\em classical resonances} which come into
play through two important, but  distinct aspects.  The first one, the
importance of  which in  this context has  already been  recognized by
Bonci  and  coworkers  \cite{BonO98PRE},   is  the  existence  of  level
crossings induced  by a resonant torus in  the semiclassical spectrum.
Alone, however,  they do not  induce any modifications, as  they arise
also in  integrable systems.  The second aspect  concerns the coupling
of  near-degenerate levels, which is related to the influence of the
destruction   of   resonant   tori   on   the   nearby   phase   space
\cite{UzeNoiMar83JCP,Ozo84JPC}.  The main purpose of this letter is to
demonstrate  that  the combination of these two aspects 
form the basis of a mechanism which, via the exponentially decaying tail  
of  the eigenstates,  controls the tunneling  between  symmetric islands.
At rather  large  values of  $\hbar$, this process  may  imply  only  one
resonance.  Deeper  in the  semiclassical regime, however,  several of
them are generally involved.

\begin{figure}[t]
\begin{center}
\leavevmode
\epsfxsize8cm
\epsfbox{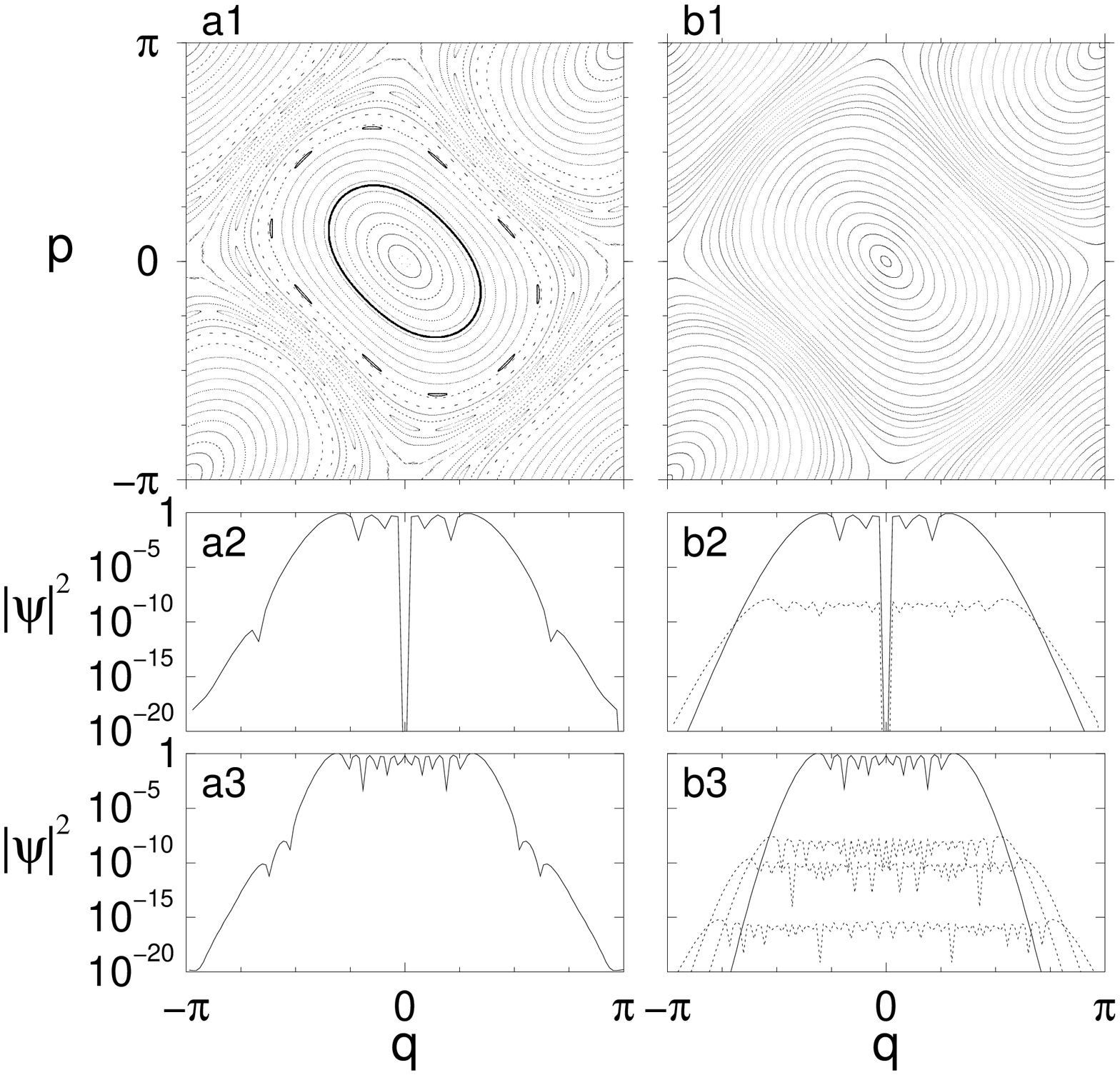}
\end{center}
\caption{
(a1) Classical phase space of the Kicked Harper at $\tau = 1$.
Boldly marked are the quantized torus with the action $I_0 = \pi / 6$ 
as well as the island chain corresponding to the 10:1 resonance.
(b1) Phase space generated by the integrable approximation $\tilde{H}$ 
(up to ${\cal O}(\tau^6)$) of the Kicked Harper map.
(a2, a3) Probability densities of Kicked Harper eigenfunctions at $\tau = 1$:
(a2) 5th excited state for $N = 2 \pi / \hbar = 66$;
(a3) 10th excited state for $N = 126$.
(b2, b3) Eigenfunctions of $\tilde{H}$ at $\tau = 1$, weighted by their 
overlap with the Kicked Harper eigenstate in (a2) and (a3), respectively:
(b2) 5th (solid line) and 15th excited state (dotted line) 
for $N = 66$;
(b3) 10th (solid line), 26th (upper dotted line), 36th (middle dotted line), 
and 50th excited state (lower dotted line) for $N = 126$.
\label{wavefunctions}}
\end{figure}

To make things more  concrete, we shall illustrate our discussion on
a specific, one-dimensional time  periodic, example, namely the kicked
Harper \cite{LebO90PRL}.   Classically,  this system is  described by
the Hamiltonian
\begin{equation}
	H \; = \; \cos p + \sum_{n = -\infty}^{\infty} \tau \, 
	\delta ( t - n \tau ) \cos q
\end{equation}
where $\tau$  represents the kick  period as well as  the perturbation
strength. Fig.~\ref{wavefunctions}(a1)  shows  the near-integrable
classical  phase space of the  corresponding map for $\tau =  1$ within 
the fundamental domain $-\pi  \leq q, p \leq  \pi$.

The quantization of the Kicked Harper map is conveniently described by
the propagator $U$ over one kick period, which is given by the product
of the two  unitary operators, $U = \exp \left(  - i\tau \hbar^{-1} \,
\cos  \hat  p \right)  \exp \left(  -  i \tau  \hbar^{-1}  \,  \cos \hat  
q \right)$,
representing  the kick  and the  free propagation  between  the kicks,
respectively.  For $\hbar  = 2 \pi / N$ with integer  $N$, the $2 \pi$
periodicity in $q$  and $p$ permits writing  the eigenfunctions $\psi$
as Bloch-like functions in both position and momentum
-- e.g., by imposing $\psi(q + 2 \pi) = \pm \psi(q)$ and 
$\hat{\psi}(p + 2 \pi) = \hat{\psi}(p)$ with $\hat{\psi}$  the
Fourier  transform  of  $\psi$.
This  effectively
reduces the eigenvalue problem $U  \psi = {\rm e}^{i \varphi} \psi$ to
the fundamental  domain, yielding, for  each particular choice  of the
periodicity  conditions, a  {\em  finite} spectrum  with $N$  discrete
eigenphases $\varphi_n$.

Fig.~\ref{wavefunctions}(a2)  shows  the   probability  density  of  a
typical eigenfunction in the near-integrable regime at $\tau = 1$, for
$\hbar = 2  \pi / N$ with $N = 66$. The  wave function
corresponds to the 5th excited state within the central regular region
around   $(q,p)=(0,0)$    and   is   localized    on   the   classical
Kolmogorov-Arnold-Moser  (KAM) torus with  the action  $I_0 =  \pi /6$
(boldly  marked   in  Fig.~\ref{wavefunctions}).   We   see  that  the
eigenfunction  exhibits  the usual  oscillatory  structure within  the
torus and decreases exponentially  beyond the caustics.  However, this
tunneling decay  is {\em not}  monotonous, but interrupted by  a local
``shoulder'' at $q \simeq \pm  0.7 \pi$.  For smaller $\hbar$ the same
description  applies  except that  several  shoulders usually appear.
See,   e.g.,   the   10th   excited   state   for   $N   =   126$   in
Fig.~\ref{wavefunctions}(a3).

\begin{figure}[t]
\begin{center}
\leavevmode
\epsfxsize7cm
\epsfbox{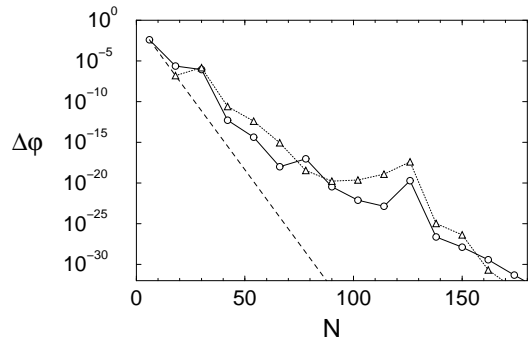}
\end{center}
\caption{  
As a function of $N = 2 \pi / \hbar$, we plot
the  difference between  the  eigenphases  of the  periodic
and  the anti-periodic  wave function
[defined by $\psi(q + 2 \pi) = \pm \psi(q)$]
for the $n$th excited state at $N = 6 ( 2 n + 1)$.
Solid line:  exact quantum  result for  the  Kicked Harper.
Dashed line: eigenphase splittings  of the corresponding states of the
integrable  approximation  $\tilde{H}$.   Dotted  line:  semiclassical
result obtained  by taking  into  account 
the effect of the 8:1, the 10:1, and the 14:1 resonance.
\label{splittings}}
\end{figure}

The  modification in  the  tunneling tail  of  the eigenfunctions  has
appreciable consequences  for   the  eigenphase   splitting  $\Delta
\varphi_n = |\varphi_n^{(+)}  - \varphi_n^{(-)}|$ between the ``periodic''
and the ``anti-periodic'' state [defined by $\psi(q + 2 \pi) = \pm \psi(q)$],
which is, in analogy to the energy difference between symmetric and
antisymmetric states in double well problems, entirely dominated by the
tunneling tails of the wave function.
Fig.~\ref{splittings}  shows,
as a function of $N = 2 \pi / \hbar$, the splittings for
the $n$th excited states at $N = 6 ( 2  n + 1 )$,
which are all localized on the 
torus with action $I_0$.   
A  clear departure   from  a  purely
exponential  decay  with  $1/\hbar$,  to be  expected  for  integrable
dynamics, is observed. 

In  order to  reveal the  cause of  this perturbation,  we  expand our
wave function   in  terms  of   eigenstates  of   the  time-independent
Hamiltonian $\tilde{H}$ that  is best adapted to the  KAM structure of
the  regular region.   
This  effective Hamiltonian can generally be expressed as an asymptotic
series in the perturbation strength $\tau$.
We use, in our case, a method introduced by Sokolov \cite{Sok86SJTMP}
and obtain
\begin{equation}
\tilde{H}  =  \cos p + \cos q - \frac{\tau}{2} \sin p \sin q 
	+ O(\tau^2)
\end{equation}
in first order in $\tau$.
Fig.~\ref{wavefunctions}(b1) shows the phase space generated by the 
integrable approximation $\tilde{H}$ up to the 6th order in $\tau$.
We see that the KAM tori of the Kicked Harper are well
reproduced within the regular region.

Calculating the  overlap $|\langle \tilde{\psi}_m |  \psi_n \rangle |$
between the  Kicked Harper eigenstate  $\psi_n$ and the  $m$th excited
eigenstate of $\tilde{H}$ identifies the structure in the
tunneling tail in  terms of contributions from the  eigenstates of the
integrable  approximation.  In  fact,  this overlap  is, as  expected,
almost  unity at  $m  = n$,  decreases  rapidly with  $|m  - n|$,  and
exhibits  relatively  pronounced local  maxima  for particular  highly
excited states  of $\tilde{H}$.  Plotting  their probability densities
[dotted  lines  in  Fig.~\ref{wavefunctions}(b2,b3)] clearly  confirms
that it is  the admixture of these particular  components which modifies
the tunneling decay of $\psi_n$.

This ``selection rule'' suggests  a resonance phenomenon as underlying
mechanism, principal  aspects of which have already  been discussed in
the  literature \cite{UzeNoiMar83JCP,Ozo84JPC}.   The  classical phase
space of the Kicked Harper  map exhibits island chains corresponding to 
nonlinear resonances between the kick and the unperturbed dynamics, two   
of which, the $10$:$1$ and the $14$:$1$ resonance, are visible in 
Fig.~\ref{wavefunctions}(a1).
A $r$:$s$ resonance, with $r$ islands, generates  in its  neighborhood a
{\em periodic perturbation} of $\tilde{H}$
and thereby couples ${\tilde \psi}_n$ to ${\tilde \psi}_{n + r k}$ with 
integer $k$, and it is precisely the combination of these couplings
that gives rise to the shoulder structures in the tunneling tail.

To demonstrate that the  above interpretation is basically correct, we
use   a   semiclassical  description   of   these  couplings   between
quasi-states to reconstruct the  wave function in the tunneling regime
and to  reproduce the behavior  of the fluctuations in  the eigenphase
splittings.   Following  the lines  of  the  derivation undertaken  by
Ozorio de Almeida \cite{Ozo84JPC},  we perform our semiclassical study
in  the action-angle  variable representation  $(I,\theta)$ associated
with the integrable dynamics  of $\tilde{H}$.  In this representation,
the Kicked Harper Hamiltonian is written as $H = H_0(I) + V(I, \theta,
t)$  with $H_0(I)  \equiv \tilde{H}(p,q)$  the  unperturbed integrable
Hamiltonian.

Let us, for the moment, assume that the phase space of the integrable
system is perturbed by only one single island chain corresponding to a
$r$:$s$  resonance.  Following  standard  secular perturbation  theory
\cite{LicLie},  we place  ourselves in  the co-rotating  frame  by the
transformation $\theta \to \hat{\theta} = \theta - 2 \pi s t / r \tau$
to the  slowly varying angle  $\hat{\theta}$.  It is then  possible to
perform a canonical  transformation towards new action-angle variables
$(I^*, \hat{\theta}^*)$ for which  the perturbation $V$ is effectively
replaced by its time-average $\bar{V}_{r:s} = (1/r\tau) \int_0^{r\tau} dt \,
V(I^*,\hat{\theta}^* + 2 \pi s t / (r \tau),t)$ over $r$ kick periods.
This yields a {\em time-independent} Hamiltonian
\begin{equation}
	H_{r:s} \; \simeq \; H_0(I^*) - \frac{2 \pi s}{r \tau} I^* 
	+ \bar{V}_{r:s}(I^*,\hat{\theta}^*), \label{Heff} 
\end{equation}
for  the effective dynamics  in  the  vicinity of  the  $r$:$s$ resonance.
The potential $\bar{V}_{r:s}$ is given by  a $2 \pi / r$ periodic function 
of $\hat{\theta}$ and can therefore be written as
\begin{equation}
	\bar{V}_{r:s}  = \sum_{k=1}^{\infty}  V^k_{r:s} \cos  (k r\hat{\theta^*} 
	+ \xi_{k}).  \label{V}
\end{equation}

In  practice,  we directly  extract  the  Fourier coefficients  $V^k_{r:s}$,
neglecting  their  action  dependence,   from  the  structure  of  the
separatrix associated with the $r$:$s$ resonance.  In the action-angle
variable        representation,       the        functional       form
$I_{sep}^{(+)}(\hat{\theta})$,  $I_{sep}^{(-)}(\hat{\theta})$  of  the
``upper''   and  the  ``lower''   separatrix  between   the  resonance
substructure and  the KAM domain  can be numerically constructed  in a
straightforward  way.  From  the knowledge  of $H_0(I)$  and  from the
condition that $H_{r:s} (I_{sep}^{(\pm)}(\hat{\theta}), \hat{\theta})$
be independent  of  $\hat{\theta}$, the coefficients $V^k_{r:s}$ are
calculated by the Fourier analysis of
\begin{equation}
	\bar{V}_{r:s}(\hat \theta) + \left[ H_0(I) - \frac{2\pi s I}{r\tau}
 	\right]_{I = I_{r:s} + \Delta I_{\rm sep}(\hat \theta)} = \mbox{const}
\end{equation}
with $I_{r:s}$ the action at the $r$:$s$ resonance and 
$\Delta  I_{\rm sep}(\hat{\theta})  =  (I_{\rm sep}^{(+)}  (\hat
\theta) - I_{\rm sep}^{(-)} (\hat{\theta}))/2$ [using 
$I_{r:s} + \Delta  I_{\rm sep}$ instead of $I_{sep}^{(\pm)}$,  
we compensate  for the  error  that is introduced  by the  discrepancy 
between  the $(I,\hat\theta)$  and the $(I^*,\hat\theta^*)$ representation].

The  Kicked   Harper  eigenstates  $\psi_n$  can   now  be  explicitly
constructed as linear combinations of the eigenstates $\tilde{\psi}_n$
of the integrable Hamiltonian within  the framework of the first order
perturbation  theory,  which  involves the  unperturbed  eigenenergies
$\tilde{E}_n$  of $\tilde{\psi}_n$  and the  matrix  elements $\langle
\tilde{\psi}_{n'} | \bar{V}_{r:s} | \tilde{\psi}_n  \rangle$   of  the
perturbation $\bar{V}_{r:s}$. These matrix elements are directly calculated
in the action-angle  variable representation $(I,\hat{\theta})$; using
Eq.~(\ref{V})   and   representing   the  unperturbed   eigenfunctions
semiclassically  as $\langle \hat{\theta}  | \tilde{\psi}_n  \rangle =
\exp(i n \hat{\theta}) / \sqrt{2 \pi}$, we obtain 
[$\sigma = {\rm sgn}(n' - n)$]
\begin{equation}
	\langle \tilde{\psi}_{n'} | \bar{V}_{r:s} | \tilde{\psi}_n \rangle 
 	=  \sum_{k=1}^{\infty} \frac{V^k_{r:s}}{2}\delta_{|n'-n|,kr} 
	\exp( i \, \sigma  \, \xi_k ).
\end{equation}
We notice that nonzero couplings arise only between states the quantum
numbers of which differ by integer multiples of $r$.  As a consequence
of this  selection rule, the perturbative expression  for the $n$th
Kicked     Harper    eigenstate     involves     only    contributions
$\tilde{\psi}_{n'}$  with $n'  -  n =  kr$  for integer  $k$, and  the
eigenphase   splittings  $\Delta   \varphi_n   =  |\varphi_n^{(+)}   -
\varphi_n^{(-)}|$ between  periodic and anti-periodic states  are to a
very  good   approximation  given   by  
\begin{eqnarray}
	\Delta  \varphi_n  &  = & 
	\sum_{k} \Gamma_{r:s}^{n,k} \Delta \tilde{\varphi}_{n+kr}
	\quad \mbox{with} \nonumber \\
	 \Gamma_{r:s}^{n,k \neq 0} & =   & \left | \frac{  V_{r:s}^k / 2  }	
	{\tilde{E}_n - \tilde{E}_{n+k r} + 2 \pi \hbar s k / \tau} \right |^2
	\label{split}
\end{eqnarray}
(and $\Gamma_{r:s}^{n,k=0} \equiv 1$).
$\Delta \tilde{\varphi}_{n} = |\tilde{\varphi}_{n}^{(+)}       -
\tilde{\varphi}_{n}^{(-)}|$ denotes the $n$th eigenphase splitting for
the  integrable  Hamiltonian $\tilde  H$  and  can be  semiclassically
computed using  standard WKB techniques for  integrable dynamics.  
In principle, the sum  on  the  r.h.s.\  of  (\ref{split})  involves 
all couplings  to the  states the  quantum  numbers of  which satisfy  the
selection  rule $n'  - n  =  kr$. In practice, it is typically  
dominated by  the
contributions with quasi-energies closest to degeneracy in the Floquet
spectrum of  the kicked system.  Such a near-degeneracy occurs  if the 
action of the $r$:$s$ resonant torus lies close to the arithmetic 
mean of the classical actions associated with ${\tilde \psi}_n$ and 
${\tilde \psi}_{n + r k}$,
in  which  case   the  energy  difference  $\tilde{E}_n  -
\tilde{E}_{n+km}$ is determined by the linear approximation of $\tilde
H$ near the resonant torus, which exhibits constant eigenphase spacing
$2\pi s/r$.

The dotted  line in Fig.~\ref{splittings}  shows the semiclassically
calculated  eigenphase  splittings of  the  states  $\psi_n$ that  are
localized on the  torus with action $I_0$.  Below $N  = 100$, only the
10:1  resonance  is taken  into  account.   Above  $N =  100$,
multiple  coupling  schemes  involving  also  the  8:1  and  14:1
resonances  become  important.
This, however, does not pose any essential difficulty.
Taking into  account the effect of each $r$:$s$ resonance separately,
we can recursively apply the mechanism described above for a single 
resonance, which amounts to generalize Eq.~(\ref{split}) to a multiple
sum over products of $\Gamma_{r:s}^{n,k}$.
This sum is again dominated by near-degenerate contributions; for
$N=126$ for instance [Fig.~\ref{wavefunctions}(a3,b3)], it is given by 
$\Delta \varphi_{10} \simeq \Gamma_{8:1}^{10,2}    \Gamma_{10:1}^{16,1}
\Gamma_{14:1}^{36,1} \Delta \tilde{\varphi}_{50}$.

We  see in Fig.~\ref{splittings} that the  agreement with  the exact  
quantum result  is fairly good,  though not  perfect. Deviations from  
the quantum  splittings occur  due  to  neglecting   the  action  
dependence  of  the  Fourier coefficients $V^k_{r:s}$  of the potential 
(\ref{V}) and sometimes also due to accidental  near-degeneracies of 
$\psi_n$ with  states that are  associated  with  the  ``opposite''  
regular region centered around $(q,p)=(\pi,\pi)$ beyond the separatrix.
Nevertheless,  the main  features  in the  fluctuations  of the  level
splittings are well reproduced by our semiclassical description.

In conclusion, we have given a simple prescription how to reproduce the
tunneling  rate fluctuations  in the  near-integrable  regime  by
the classical structure of  a few important resonances.  The effect of
a $r$:$s$ resonance  on  the  tunneling  process  of  a  regular  
eigenstate  is twofold. 
On  the one  hand, the periodic  modulation of the  KAM torus
structure due to the resonance induces appreciable couplings to higher
states that  are selected  by  the periodicity  $r$ of  the
island chain. 
These  couplings become, on the  other hand, substantially
enhanced at near-degeneracies in the  quasi-energy spectrum. We
have  explicitly  incorporated   this  mechanism  in  a perturbative 
semiclassical framework which, through simple iteration, is
generalized to multiple coupling schemes involving several resonances.
As underlying picture, we obtain that each $r$:$s$ resonance allows the  
wave function to ``hop'' from one side of the island chain to the other,
with the quantum number changing by integer multiples of $r$.
For large $\hbar$ only one or few resonances with relatively small $r$ 
come into play, whereas for small $\hbar$ the tunnel
coupling proceeds via a succession of hops over several adjacent island
chains, rather than by one single, but  longer, classically forbidden
event. Comparison with the quantum tunneling rates shows good agreement, 
and confirms in particular also that contributions from the chaotic part 
of phase space do not play a role in the near-integrable regime.

Though exemplified only within the Kicked Harper model, we expect that the
resonance-assisted coupling  phenomenon discussed here  represents the
fundamental tunneling mechanism for general near-integrable systems.
An open  and interesting question is to which extent  the role  of the  
resonances  retains its  importance also  in strongly mixed regular-chaotic 
systems where a macroscopic fraction of quantum states is associated with 
chaotic domains.  We believe that at least the  ``regular'' contribution to  
chaos-assisted tunneling, i.e., the coupling from the torus to the chaos
border,  is  now  amenable  to straightforward  reproduction  via  the
identification of the major resonances inside the regular domain.

We  thank  E.~Bogomoly, O.~Bohigas,  P.~Leboeuf,  S.~Tomsovic, and 
A.~M.~Ozorio de Almeida  for  helpful  and  inspiring  discussions.   
PS  acknowledges financial support from Alexander von Humboldt-Stiftung 
and DFG. The LPTMS is an ``Unit\'e de recherche de l'Universit\'e Paris 11
associ\'ee au C.N.R.S.''

\end{document}